\documentclass[prd,aps,floats,onecolumn,twoside,preprintnumbers,
superscriptaddress,floatfix,nofootinbib,showpacs]{revtex4}
\usepackage{graphicx,pstricks,rotating,amsmath}
\usepackage{slashed}

\begin{document}

\title{The $\theta$-exact Seiberg-Witten maps at the $e^3$ order}

\author{Josip Trampeti\'{c}}
\affiliation{Max-Planck-Institut f\"ur Physik,
	(Werner-Heisenberg-Institut),
  	 F\"ohringer Ring 6, D-80805 M\"unchen, Germany}
\email{trampeti@mppmu.mpg.de}
\affiliation{Institute Rudjer Bo\v{s}kovi\'{c}, Division of Theoretical Physics, Bijeni\v{c}ka 54 10000 Zagreb, Croatia}
\email{josipt@rex.irb.hr, youjiangyang@gmail.com}
\author{Jiangyang You}
\affiliation{Institute Rudjer Bo\v{s}kovi\'{c}, Division of Theoretical Physics, Bijeni\v{c}ka 54 10000 Zagreb, Croatia}

\begin{abstract}
We study two distinct $\theta$-exact Seiberg-Witten (SW) map expansions, (I) and (II) respectively, up to $e^3$ order for the gauge parameter, gauge field and the gauge field strengths of the noncommutative $\rm U_\star(1)$ gauge theory on the Moyal space. We derive explicitly the closed form expression for the SW map ambiguity between the two and observe the emergence of several new totally commutative generalized star products. We also identify the additional gauge freedoms within each of the $e^3$ order field strength expansions and define corresponding sets of deformation/ratio/weight parameters, $(\kappa$, $\kappa_{i})$ and $(\kappa$, $\kappa'_{i})$, for these two SW maps respectively.
\end{abstract} 



\maketitle

\section{Introduction}
The $\theta$-exact Seiberg-Witten (SW) map is an old and new subject in the noncommutative (NC) gauge field theory (NCGFT) on the Moyal space. Some results emerged immediately after the map itself~\cite{Seiberg:1999vs} was discovered~\cite{Jurco:2000fb,Madore:2000en,Jurco:2000ja,Schupp:2001we,Jurco:2001kp,Mehen:2000vs,Liu:2000ad,Liu:2000mja,Okawa:2001mv,Szabo:2001kg,Jurco:2001my,Jurco:2001rq,Brace:2001fj,Brace:2001rd,Cerchiai:2002ss,Calmet:2001na,Behr:2002wx,Aschieri:2002mc,Schupp:2002up,Minkowski:2003jg,Barnich:2002pb,Barnich:2003wq,Banerjee:2003vc,Banerjee:2004rs,Trampetic:2007hx,Horvat:2011qn}. Applications to the perturbative noncommutative quantum field theories (NCQFT) started several years later. Till now it has been shown to be of great value for developing nontrivial variants of the NCQFT from both theoretical and phenomenological perspectives~\cite{Martin:1999aq,Bichl:2001cq,Martin:2002nr,Brandt:2003fx,Buric:2006wm,Buric:2007qx,Schupp:2008fs,Blaschke:2009aw,Szabo:2009tn,Horvat:2010km,Horvat:2011bs,Horvat:2011qg,Wang:2012ye,Horvat:2013rga,Wang:2013iga,Horvat:2012vn}. Yet most of them are restricted to the first/$e^2$ order of the $\theta$-exact expansion only due to the complicatedness of (obtaining) the second/$e^3$ order expansion.\footnote{Generically the ordering employed in this article is the formal power of field operators or the homogeneity in the fields~\cite{Barnich:2003wq}. The equivalent notation of coupling constant ordering is a consequence of the so-called $\rm U_\star(1)$ charge quantization issue and its resolution within the SW map approach~\cite{Schupp:2001we,Horvat:2011qn}. In this resolution~\cite{Schupp:2001we,Horvat:2011qn} the commutative field in the SW map expansion for a $\rm U_\star(1)$ theory are bundled with the commutative charge $Qe$ in order to normalize the NC field to the quantized charges $0,\pm 1$, which in turn induces the equivalence between the field operator and coupling constant power ordering. We use the name coupling constant ordering for its easier visibility in (front of) the equations.} Recently a systematic construction of the $\theta$-exact SW map expansion with respect to the powers of coupling constant $e$  for abitrary gauge field theories on Moyal space was proposed in \cite{Martin:2012aw}. (See also~\cite{Martin:2015nna} for its latest hybrid SW map extension.) This could trigger many further applications in the near future.

The main aim of this paper is to continue the aforementioned important progress on the $\theta$-exact SW map expansion~\cite{Martin:2012aw}. Here we focus on two topics: First, the $e^3$ order SW map expansion for $\rm U_\star(1)$ gauge field obtained via the method in~\cite{Martin:2012aw} (denoted as SW map (I)) appears to be different from the early results~\cite{Mehen:2000vs,Schupp:2008fs} (SW map (II)), which is not really surprising since it is long recognized that SW map is far from unique when defined as the map between noncommutative and commutative fields that preserves the smooth commutative limit and satisfies the consistency condition relations ~\cite{Asakawa:1999cu,Jurco:2001my,Jurco:2001rq,Brace:2001fj,Brace:2001rd,Cerchiai:2002ss,Barnich:2002pb,Barnich:2003wq}:
\begin{gather}
\delta_\Lambda A_\mu \equiv \partial_\mu\Lambda+ i[\Lambda \stackrel{\star}{,}A_\mu] =\delta_\lambda A_\mu[a_\mu]\,,
\label{deltaA}\\
\delta_\Lambda F_{\mu\nu} \equiv i[\Lambda \stackrel{\star}{,}F_{\mu\nu}]=\delta_\lambda F_{\mu\nu}[a_\mu]\,,
\label{deltaF}\\
\Lambda[[\lambda_1,\lambda_2],a_\mu]
=[\Lambda[\lambda_1,a_\mu]\stackrel{\star}{,}
\Lambda[\lambda_2,a_\mu]]
+i\delta_{\lambda_1}\Lambda[\lambda_2,a_\mu]-i\delta_{\lambda_2}
\Lambda[\lambda_1,a_\mu]\,.
\label{SWrecurs}
\end{gather}
The Moyal star($\star$)-product used in the above is defined as usual.
Besides possible connection by gauge transformation, there can be plenty of generic ambiguity/freedom/redundancy between two different gauge field SW maps. Still it gives rise to the question to which extent exactly are the two maps (I) and (II) different from each other. The ambiguity between the NC gauge field expansions can be characterized by looking at the composition of the first SW map expansion $\{\Lambda_1(a_\mu,\lambda),A_{1_\mu}(a_\mu)...\}$ and by the inverse $\{\lambda_2(A_\mu,\Lambda),a_{2_\mu}(A_\mu)...\}$ of the second SW map expansion $\{\Lambda_2(a_\mu,\lambda),A_{2_\mu}(a_\mu)...\}$~\cite{Barnich:2003wq}. Consistency conditions then lead to the following equality in the case of the NC $\rm U_\star(1)$ gauge theory
\begin{equation}
\partial_\mu\lambda_2\big(A_{1_\mu}(a_\mu),\Lambda_1(a_\mu,\lambda)\big)=\delta_\lambda a_{2_\mu}\big(A_{1_\mu}(a_\mu)\big).
\label{1.1}
\end{equation}
It was further pointed out~\cite{Barnich:2002pb,Barnich:2003wq} that such composition bears the following general form
\begin{gather}
a_{2_\mu}\big(A_{1_\mu}(a_\mu)\big)=a_\mu+X_\mu(a_\mu)+\partial_\mu Y(a_\mu),
\label{1.2}\\
\lambda_2\big(A_{1_\mu}(a_\mu),\Lambda_1(a_\mu,\lambda)\big)=\lambda+\delta_\lambda Y(a_\mu),
\label{1.3}
\end{gather}
where $\delta_\lambda X_\mu(a_\mu)=0$. Thus, the procedure to thoroughly distinguish these two SW maps would be to determine $X_\mu(a_\mu)$ and $Y(a_\mu)$ explicitly.

Second, a large class of SW map ambiguities (field redefinitions) was constructed in the past by iteratively adding commutative gauge covariant terms with free coefficients at the $n$-th power of $\theta^{\mu\nu}$ to the known solution $A^{\theta^n}_\mu$ up to the same order~\cite{Asakawa:1999cu,Bichl:2001cq}. It is found that such redefinitions contribute to the field strength in the action which can help to cancel certain divergences in the perturbative quantum loop computations~\cite{Buric:2006wm,Latas:2007eu,Buric:2007ix,Buric:2010wd,Martin:2009sg,Martin:2009vg}. These findings made such method highly favorable in the $\theta$-expanded studies, yet it remains an open question how to generalize this procedure to the $\theta$-exact approach since the iteration is based solely on the powers of $\theta^{ij}$ and consequently gives no hint about how to re-sum over all orders of $\theta$. Recently we observe that the $e^2$ order $\theta$-exact expansion of the field strength for $\rm U_\star(1)$ gauge theory possesses a freedom by itself~\cite{Trampetic:2014dea}: It contains a term invariant under the commutative gauge transformation. For this reason this term does not contribute to the consistency relation at $e^2$ order. One can then freely vary the ratio of this term with respect to the other term. The resulting deformed field strength operator still allows the usual quadratic action to be gauge invariant up to $e^2$ order. The $e^2$ order ratio coefficient can be shown to be the inverse of the $\theta^1$-order iteration induced coefficient reported in the early works. This fact motivates us to consider such ratio parameter(s) a possible substitute for the iteration induced coefficient(s) in the $\theta$-exact approach.

In this article we study both topics mentioned above at the $e^3$ order. We first compare, up to the cubic order of the coupling constant $e$, two distinct $\theta$-exact Seiberg-Witten map expansions for the NC $\rm U_\star(1)$ gauge theory: One obtained from Seiberg-Witten differential equation (I), and the other by inverting an early $\theta$-exact inverted SW map solution (II). We give a closed form expression for the SW map ambiguity between these two maps and show that this ambiguity/freedom could contribute to the field strength. We then extend the procedure in~\cite{Trampetic:2014dea} to each of the $e^3$ order field strength SW map expansion, identify the gauge invariant parts inside each of the expansions and assign the corresponding ratio parameters. With the help from the new generalized star products found when studying the gauge field ambiguity, we are able to explicitly express each of the gauge invariant parts in terms of the commutative field strength too.

The paper is structured as follows: In the second section we describe both $\theta$-exact Seiberg-Witten map expansion solutions up to the $e^3$ order. The SW map ambiguity between these two SW maps is given explicitly in Section III, where we also demonstrate the emergence of several new totally commutative generalized star products within the expressions. Section IV is devoted to the freedoms within each of the $e^3$ order $\theta$-exact gauge field strength expansions respectively. Discussions and conclusions then follow. In this article the capital letters denote noncommutative objects, and the small letters denote commutative objects.

\section{Two different $\theta$-exact Seiberg-Witten expansions up to the $e^3$  order}
\subsection{Seiberg-Witten map (I)}
The first powerful method (I) to obtain $\theta$-exact SW map expansion for noncommutative gauge theories on Moyal space is by solving the SW differential equations \cite{Seiberg:1999vs,Brace:2001fj,Brace:2001rd,Cerchiai:2002ss,Barnich:2002pb,Barnich:2003wq,Martin:2012aw,Martin:2015nna}. For the NC gauge parameter ($\Lambda$), the NC gauge field $(A_\mu)$, and the NC gauge field strength $(F_{\mu\nu})$  of the $\rm U_\star(1)$ gauge theory, these equations read
\begin{gather}
\frac{d}{dt} \Lambda (x)=-\frac{1}{4}\theta^{ij}\Big\{A_i\stackrel{\star_t}{,}\partial_j\Lambda\Big\},
\label{2.1}\\
\frac{d}{dt} A_\mu(x)=\frac{1}{4}\theta^{ij}\Big\{A_i\stackrel{\star_t}{,}\partial_j A_\mu+F_{j\mu}\Big\},
\label{2.2}\\
\frac{d}{dt} F_{\mu\nu}(x)=\frac{1}{4}\theta^{ij}\bigg[\Big\{F_{\mu i}\stackrel{\star_t}{,}F_{\nu j}\Big\}-\Big\{A_i\stackrel{\star_t}{,}\left(D^{\star_t}_j +\partial_j \right)F_{\mu\nu}\Big\}\bigg],
\label{2.3}
\end{gather}
where the Moyal $\star$-product with an additional parameter $t$ is defined as:
\begin{equation}
(\phi\star_t \psi)(x)=
e^{\frac{i}{2}t\theta^{ij}{{\partial}^\eta_i}\,
\,{{\partial}^\xi_j}} \phi(x+\eta)\psi(y+\xi)\big|_{\eta,\xi\to0}
\equiv\phi(x) e^{\frac{i}{2}\overleftarrow{{\partial}_i}\,
t\theta^{ij}\,\overrightarrow{{\partial}_j}} \psi(x).
\label{f*g}
\end{equation}
Note that in the rest of the article this parameter $t$ will be absorbed into the definition of $\theta^{ij}$ when not needed. The noncommutative covariant derivative is defined in the following way $D^{\star_t}_j =\partial_j - i [A_j\stackrel{\star_t}{,}{\;\;}]$. By imposing initial conditions~\cite{Horvat:2011qg}
\begin{gather}
\Lambda_{_{\rm (I)}}(x)=e\lambda+\mathcal O(e^2)\,,\;\;
A_{\mu_{{\rm (I)}}}(x)=e a_\mu+\mathcal O(e^2)\,,
\label{2.41}
\end{gather}
one can easily solve equations (\ref{2.1}) and (\ref{2.2}) at the $e^2$ order and obtain the following solutions
\begin{gather}
\Lambda_{_{\rm (I)}}(x)=e\lambda-\frac{e^2}{2}\theta^{ij}a_i\star_{2_t}\partial_j\lambda+\mathcal O(e^3)\,,\;\;
A_{\mu_{{\rm (I)}}}(x)=e a_\mu-\frac{e^2}{2}\theta^{ij}a_i\star_{2_t}(\partial_j a_\mu+f_{j\mu})+\mathcal O(e^3).
\label{2.5}
\end{gather}
Then, the next order of the SW differential equations can be written down recursively
\begin{gather}
\begin{split}
\frac{d}{dt} \Lambda^{e^3}(x)=\frac{e^3}{8}\theta^{ij}\theta^{kl}\bigg[\Big\{a_i\stackrel{\star_t}{,}\partial_j\left(a_k\star_{2_t}\partial_l\lambda\right)\Big\}
+\Big\{a_k\star_{2_t}(\partial_l a_i+f_{li})\stackrel{\star_t}{,}\partial_j\lambda\Big\}\bigg],
\end{split}
\label{2.6}\\
\begin{split}
\frac{d}{dt} A^{e^3}_\mu(x)=\frac{e^3}{8}\theta^{ij}\theta^{kl}&\bigg[\Big\{a_i\stackrel{\star_t}{,}\Big(\partial_j\big(a_k\star_{2_t}(\partial_l a_\mu+f_{l\mu})\big)-2(f_{jk}\star_{2_t}f_{\mu l}-a_i \star_{2_t}\partial_l f_{j\mu})\Big)\Big\}
\\&+\Big\{\big(a_k\star_{2_t}(\partial_l a_i+f_{li})\big)\stackrel{\star_t}{,}(\partial_j a_\mu+f_{j\mu})\Big\}\bigg].
\end{split}
\label{2.7}
\end{gather}
The $\star_{2_t}$-product is defined analogous to the $\star_t$-product, i.e.
\begin{equation}
\phi(x)\star_{2_t} \psi(x)=\frac{\sin \big(t\frac{\partial_1\theta \partial_2}{2}\big)}{\big(\frac{\partial_1\theta
\partial_2}{2}\big)}\phi(x_1)\psi(x_2)\bigg|_{x_1=x_2=x}.
\label{f*2tg}
\end{equation}
The fact that denominator is not scaled by $t$ is crucial in solving the higher order SW differential equations \cite{Martin:2012aw}. On the other hand the $t^n$ ordering still remains the same as $\theta^n$ ordering because of the extra $\theta^{ij}$ outside the generalized star product(s).

Since equations (\ref{2.6}) and (\ref{2.7}), involve only $\{f\stackrel{\star_t}{,}(g\star_{2_t}h)\}$ type terms, in accord with technique from~\cite{Martin:2012aw} one can immediately introduce a new generalized $\star_{3'}$-product
\begin{equation}
\begin{split}
[fgh]_{\star_{3'}}=\int\limits_{0}^{t}\,d{t'}\,\{f\stackrel{\star_{t'}}{,}(g\star_{2_{t'}}h)\}=&\cdot\Bigg(\frac{\cos \left[t\left(\frac{\partial_f\theta\partial_g}{2}+\frac{\partial_f\theta\partial_h}{2}-\frac{\partial_g\theta\partial_h}{2}\right)\
\right]-1}{\left(\frac{\partial_f\theta\partial_g}{2}+\frac{\partial_f\theta\partial_h}{2}-\frac{\partial_g\theta\partial_h}{2}\right)\left(\frac{\partial_g\theta\partial_h}{2}\right)}
\\&-\frac{\cos \left[t\left(\frac{\partial_f \theta\partial_g}{2}+\frac{\partial_f \theta\partial_h}{2}+\frac{\partial_g \theta\partial_h}{2}\right)\right]-1}{\left(\frac{\partial_f\theta\partial_g}{2}+\frac{\partial_f\theta\partial_h}{2}+\frac{\partial_g\theta\partial_h}{2}\right)\left(\frac{\partial_g\theta\partial_h}{2}\right)}\Bigg)f\otimes g\otimes h\,,
\label{fgh3'}
\end{split}
\end{equation}
for a universal expression of the $e^3$ order expansion. In this notation we find following $\theta$-exact solutions for the SW differential equations up to the $e^3$ order:
\begin{gather}
\begin{split}
\Lambda_{_{\rm (I)}}(x)=e\lambda-\frac{e^2}{2}\theta^{ij}a_i\star_2\partial_j\lambda+\frac{e^3}{8}\theta^{ij}\theta^{kl}\Big[a_i\partial_j(a_k\partial_l\lambda) -\partial_i\lambda a_k(\partial_l a_j+f_{lj})\Big]_{\star_{3'}}+\mathcal O\left(e^4\right),
\end{split}
\label{2.9}\\
\begin{split}
A_{\mu_{\rm (I)}}(x)=&e a_\mu-\frac{e^2}{2}\theta^{ij}a_i\star_2\left(\partial_j a_\mu+f_{j\mu}\right)+\frac{e^3}{8}\theta^{ij}\theta^{kl}\bigg(\Big[a_i\partial_j\big(a_k(\partial_l a_\mu+f_{l\mu})\big)\Big]_{\star_{3'}}
\\&-2\Big[a_i(f_{jk}f_{l\mu}-a_k\partial_l f_{j\mu})\Big]_{\star_{3'}}
+\Big[(\partial_j a_\mu+f_{j\mu})a_k(\partial_l a_i+f_{li})\Big]_{\star_{3'}}\bigg)+\mathcal O\left(e^4\right).
\end{split}
\label{2.10}
\end{gather}

\subsection{Seiberg-Witten map (II)}
Another type (II) of $\theta$-exact Seiberg-Witten map expansion~\cite{Schupp:2008fs} was obtained by inverting the solutions from \cite{Mehen:2000vs}, the explicit expansion for $A_\mu$ and $\Lambda$ up to $e^3$ order is as follows
\begin{gather}
\begin{split}
\Lambda_{_{\rm (II)}}(x)&=e\lambda-\frac{e^2}{2}\theta^{ij}a_i\star_2\partial_j\lambda+\frac{e^3}{2}\theta^{ij}\theta^{kl}\bigg[\frac{1}{2}\big(a_k\star_2(\partial_l a_i+f_{li})\big)\star_2\partial_j\lambda
\\&+\frac{1}{2}a_i\star_2\partial_j(a_k\star_2\partial_l\lambda)\bigg]-\frac{e^3}{2}\theta^{ij}\theta^{kl}\Big[\partial_k\partial_i\lambda a_j a_l+\partial_k\lambda a_i\partial_l a_j\Big]_{\star_3}+\mathcal O(e^4),
\label{2.11}
\end{split}
\\
\begin{split}
A_{\mu_{\rm (II)}}(x)&=ea_\mu-\frac{e^2}{2}\theta^{ij}a_i\star_2(\partial_j
a_\mu+f_{j\mu})
\\&+\frac{e^3}{2}\theta^{ij}\theta^{kl}\bigg[\frac{1}{2}\big(a_k\star_2(\partial_l
a_i+f_{li})\big)\star_2(\partial_j
a_\mu+f_{j\mu})+a_i\star_2\Big(\partial_j \big(a_k\star_2(\partial_l
a_\mu+f_{l\mu})\big)
\\&-\frac{1}{2}\partial_\mu \big(a_k\star_2(\partial_l
a_j+f_{lj})\big)\Big)
-\frac{1}{2}a_i\star_2(\partial_k
a_j\star_2\partial_l
a_\mu) \bigg]
\\&+\frac{e^3}{2}\theta^{ij}\theta^{kl}\Big[a_i\partial_k
a_\mu(\partial_j a_l+f_{jl})-\partial_k\partial_i a_\mu a_j a_l
-2\partial_k a_i\partial_\mu a_j a_l\Big]_{\star_3}+\mathcal O(e^4).
\label{2.12}
\end{split}
\end{gather}
Clearly this gives the same the $e^2$ order solution as in the equations (\ref{2.9}) and (\ref{2.10}) of the SW map (I), however the $e^3$ order starts to show difference. The totally commutative $\star_3$-product~\cite{Mehen:2000vs} is defined as follows
\begin{equation}
[f(x)g(x)h(x)]_{\star_3}=\Bigg(\frac{\sin(\frac{\partial_2\theta
\partial_3}{2})\sin(\frac{\partial_1\theta(\partial_2+\partial_3)}{2})}
{\frac{(\partial_1+\partial_2)\theta \partial_3}{2}
\frac{\partial_1\theta(\partial_2+\partial_3)}{2}}
+\{1\leftrightarrow 2\}\,\Bigg)f(x_1)g(x_2)h(x_3)\Bigg|_{x_i=x}\,.
\label{star3}
\end{equation}

The generalized star products $\star_2$, $\star_3$ and $\star_{3'}$ are connected with star commutators by the following relations
\begin{gather}
[f(x)\stackrel{\star}{,}g(x)]=i\theta^{ij}\partial_i f(x)\star_2\partial_j g(x),
\label{fe2}
\\
f(x)\star_2[g(x)\stackrel{\star}{,}h(x)]+g(x)\star_2[f(x)\stackrel{\star}{,}h(x)]=i\theta^{ij}[\partial_i f(x)g(x)\partial_j h(x)+f(x)\partial_i g(x)\partial_j h(x)]_{\star_3},
\label{fe3}
\\
[f(x)\stackrel{\star}{,}g(x)\star_2 h(x)]=\frac{i}{2}[h(x)\partial_i f(x)\partial_j g(x)+g(x)\partial_i f(x)\partial_j h(x)+\partial_i f(x)\partial_j g(x)h(x)+\partial_i f(x)g(x)\partial_j h(x)]_{\star_{3'}}.
\label{fe3p}
\end{gather}
Here we see that there are extra derivatives in the generalized star product formula for the star commutators, which provides the opportunity to "integral over" the infinitesimal transformation $\partial_i\lambda\to a_i$. Note also the difference between $\star_3$ and $\star_{3'}$ products: $\star_{3'}$ helps realizing the $[f(x)\stackrel{\star}{,}g(x)\star_2 h(x)]$ structure, which takes place in the infinitesimal commutative gauge transformation of the SW map expansion of the NC $\rm U_\star(1)$ gauge field in terms of commutative $\rm U(1)$ gauge field; while $\star_3$ realizes the $f(x)\star_2\left[g(x)\stackrel{\star}{,}h(x)\right]$ which is typical in the the inverse SW map expansion of the commutative $\rm U(1)$ gauge field in terms of the NC $\rm U_\star(1)$ gauge field~\cite{Mehen:2000vs}.

\section{The $\theta$-exact Seiberg-Witten map ambiguity at the $e^3$ order}

The two $e^3$ order Seiberg-Witten map expansions presented in the last section look quite different although they bear similarity to the certain degree: They both start at the order $\theta^2$ and bear similar tensor structure. It is not hard to show that they are indeed not equal to each other by, for example, inspecting the $\theta^2$ and $\theta^4$ order expansions of each solution. Therefore certain ambiguity structure should exist between those two solutions. In this section we compare these two $\theta$-exact SW maps up to the $e^3$ order given in section II in detail. Following the arguments in~\cite{Barnich:2002pb,Barnich:2003wq}, we consider the composition of one of the SW map and the inverse of the other. Now, since SW map (II) were derived from a $\theta$-exact inverse SW map expansion in \cite{Mehen:2000vs}, we choose the original inverse map of (II) for the ambiguity analysis outlined in the introduction. This inverse SW map expansion is as follows
\begin{gather}
\begin{split}
\lambda_{_{\rm (II)}}(A_\mu,\Lambda)=\Lambda+\frac{1}{2}\theta^{ij}\Big(A_i\star_2\partial_j\Lambda+\theta^{kl}\big[\partial_i\partial_k\Lambda A_j A_l+\partial_i\Lambda A_k\partial_j A_l\big]_{\star_3}\Big)+\mathcal O(A^3)\Lambda,
\end{split}
\label{3.1}\\
\begin{split}
a_{\mu_{\rm (II)}}(A_\mu)=&A_\mu+\frac{1}{2}\theta^{ij}\Big(A_i\star_2\left(\partial_j A_\mu+F_{j\mu}\right)
+\theta^{kl}\big[-A_i\partial_k A_\mu(\partial_j A_l+F_{jl})+\partial_i\partial_k A_\mu A_j A_l+\partial_k A_i\partial_\mu A_j A_l\big]_{\star_3}\Big)+\mathcal O(A^4).
\end{split}
\label{3.2}
\end{gather}
Now expanding the compositions $\lambda_{_{\rm (II)}}\big(A_{\mu_{\rm (I)}}(a_\mu),\Lambda_{_{\rm (I)}}(a_\mu,\lambda)\big)$ and $a_{\mu_{\rm (II)}}\big(A_{\mu_{\rm (I)}}(a_\mu)\big)$ up to the $e^3$ order, we find
\begin{gather}
\lambda_{_{\rm (II)}}\big(A_{\mu_{\rm (I)}}(a_\mu),\Lambda_{_{\rm (I)}}(a_\mu,\lambda)\big)=e\lambda(x)+\Lambda^{e^3}_{_{\rm (I)}}(x)-\Lambda^{e^3}_{_{\rm (II)}}(x)+\mathcal O(e^4),
\label{3.3}\\
a_{\mu_{\rm (II)}}\big(A_{\mu_{\rm (I)}}(a_\mu)\big)=ea_\mu(x)+A^{e^3}_{\mu_{\rm (I)}}(x)-A^{e^3}_{\mu_{\rm (II)}}(x)+\mathcal O(e^4).
\label{3.4}
\end{gather}
Note that the $e^2$ order vanishes as expected. Equations \eqref{1.2} and \eqref{1.3} in the introduction then indicate that the (I) minus (II) differences at the $e^3$ order should bear following expressions:
\begin{equation}
A^{e^3}_{\mu_{\rm (I)}}(x)-A^{e^3}_{\mu_{\rm (II)}}(x)=X^{e^3}_\mu(x)+\partial_\mu Y^{e^3}(x)\,,\;\;
\Lambda^{e^3}_{_{\rm (I)}}(x)-\Lambda^{e^3}_{_{\rm (II)}}(x)=\delta_\lambda Y^{e^3}(x).
\label{3.6}
\end{equation}
In order to find solution for the explicit forms of $X^{e^3}_\mu(x)$ and $Y^{e^3}(x)$, we first Fourier transform $A^{e^3}_\mu(x)$ into a momentum space quantity $\tilde A^{e^3}_\mu(p,q,k)$:
\begin{equation}
\begin{split}
\tilde A^{e^3}_\mu(p,q,k)=
-\frac{e^3}{8}&\Big[\tilde a_\mu(k) \Big(\big(\tilde a(p)\theta q\big)\big(\tilde a(q)\theta k\big)M_1+\big(\tilde a(p)\theta\tilde a(q)\big)(q\theta k)M_2
+\big(\tilde a(p)\theta k \big)\big(\tilde a(q)\theta k \big)M_3 \Big)
\\&+k_\mu\Big(\big(\tilde a(p)\theta q\big)\big(\tilde a(q)\theta \tilde a(k)\big)M_4
+\big(\tilde a(p)\theta\tilde a(q)\big)\big(q\theta \tilde a(k)\big)M_5+\big(\tilde a(p)\theta\tilde a(k)\big)\big(\tilde a(q)\theta k \big)M_6\Big)\Big].
\label{tildeA}
\end{split}
\end{equation}
Then from equations \eqref{2.9} to \eqref{2.12} we read out the coefficients $M_i$s for the SW maps ${\rm (I)}$ and ${\rm (II)}$ respectively:
\begin{equation}
\begin{split}
&M_{1_{\rm (I)}}=4f_{\star_{3'}}\left(\frac{p\theta q}{2},\frac{p\theta k}{2},\frac{q\theta k}{2}\right)+4f_{\star_{3'}}\left(-\frac{q\theta k}{2},-\frac{p\theta k}{2},-\frac{p\theta q}{2}\right)=-2M_{2_{\rm (I)}},
\\&
M_{3_{\rm (I)}}=4f_{\star_{3'}}\left(\frac{p\theta q}{2},\frac{p\theta k}{2},\frac{q\theta k}{2}\right),
\\&
M_{4_{\rm (I)}}=-3f_{\star_{3'}}\left(\frac{p\theta q}{2},\frac{p\theta k}{2},\frac{q\theta k}{2}\right)-2f_{\star_{3'}}\left(-\frac{q\theta k}{2},-\frac{p\theta k}{2},-\frac{p\theta q}{2}\right),
\\&
M_{5_{\rm (I)}}=-2f_{\star_{3'}}\left(\frac{p\theta q}{2},\frac{p\theta k}{2},\frac{q\theta k}{2}\right)-f_{\star_{3'}}\left(-\frac{q\theta k}{2},-\frac{p\theta k}{2},-\frac{p\theta q}{2}\right),\\
\\&
M_{6_{\rm (I)}}=2f_{\star_{3'}}\left(\frac{p\theta q}{2},\frac{p\theta k}{2},\frac{q\theta k}{2}\right)+f_{\star_{3'}}\left(\frac{q\theta k}{2},-\frac{p\theta q}{2},-\frac{p\theta k}{2}\right),
\label{SWM1-6}
\end{split}
\end{equation}
and
\begin{equation}
\begin{split}
M_{1_{\rm (II)}}&=8f_{\star_2}\left(\frac{p\theta q}{2}\right)f_{\star_2}\left(\frac{(p+q)\theta k}{2}\right)
+8f_{\star_2}\left(\frac{q\theta k}{2}\right)f_{\star_2}\left(\frac{p\theta (q+k)}{2}\right)
-8f_{\star_3}\left(\frac{p\theta q}{2},\frac{p\theta k}{2},\frac{q\theta k}{2}\right)=-2M_{2_{\rm (II)}},
\\
M_{3_{\rm (II)}}&=8f_{\star_2}\left(\frac{q\theta k}{2}\right)f_{\star_2}\left(\frac{p\theta (q+k)}{2}\right)-4f_{\star_3}\left(\frac{p\theta q}{2},\frac{p\theta k}{2},\frac{q\theta k}{2}\right),
\\
M_{4_{\rm (II)}}&=-4f_{\star_2}\left(\frac{p\theta q}{2}\right)f_{\star_2}\left(\frac{(p+q)\theta k}{2}\right)
-6f_{\star_2}\left(\frac{q\theta k}{2}\right)f_{\star_2}\left(\frac{p\theta (q+k)}{2}\right)
+8f_{\star_3}\left(\frac{p\theta q}{2},\frac{p\theta k}{2},\frac{q\theta k}{2}\right),
\\
M_{5_{\rm (II)}}&=2f_{\star_2}\left(\frac{p\theta q}{2}\right)f_{\star_2}\left(\frac{(p+q)\theta k}{2}\right)+4f_{\star_2}\left(\frac{q\theta k}{2}\right)f_{\star_2}\left(\frac{p\theta (q+k)}{2}\right),
\\
M_{6_{\rm (II)}}&=-4f_{\star_2}\left(\frac{q\theta k}{2}\right)f_{\star_2}\left(\frac{p\theta (q+k)}{2}\right)-2f_{\star_2}\left(\frac{p\theta k}{2}\right)f_{\star_2}\left(\frac{q\theta (p+k)}{2}\right).
\label{NSWM1-6}
\end{split}
\end{equation}
The functions $f_{\star_2}(a)$, $f_{\star_{3}}(a,b,c)$, and $f_{\star_{3'}}(a,b,c)$ are defined as below:
\begin{gather}
f_{\star_2}\left(a\right)=\frac{\sin a}{a}\,,
\label{f2}\\
f_{\star_3}\left(a,b,c\right)=\frac{\sin b\,\sin(a+b)}{(a+b)(b+c)}+\frac{\sin c\,\sin(a-c)}{(a-c)(b+c)}\,,
\label{f3}\\
f_{\star_{3'}}\left(a,b,c\right)=\frac{\cos(a+b-c)-1}{(a+b-c)c}-\frac{\cos(a+b+c)-1}{(a+b+c)c}\,.
\label{f3'}
\end{gather}
From equations (\ref{SWM1-6}) and (\ref{NSWM1-6}) we observe that under the permutation $q\leftrightarrow k$:
\begin{equation}
M_{5_{\rm (I,II)}}=-M_{6_{\rm (I,II)}}(q\leftrightarrow k)\,,
\label{M56}
\end{equation}
which indicate that the $M_{5_{\rm (I,II)}}$ and $M_{6_{\rm (I,II)}}$ (and part of the $M_{4_{\rm (I,II)}}$)  contributions could be made equivalent to two appropriate infinitesimal NC gauge transformation(s) respectively
\begin{equation}
\begin{split}
A'^{e^3}_{\mu_{\rm (I,II)}}(x)&=A^{e^3}_{\mu_{\rm (I,II)}}(x)+\delta_{\Xi^{e^3}_{\rm (I,II)}}A^{e^3}_{\mu_{\rm (I,II)}}(x)
=A^{e^3}_{\mu_{\rm (I,II)}}(x)+\partial_\mu\Xi^{e^3}_{\rm (I,II)}(x)+\mathcal(O)(e^4).
\end{split}
\label{3.14}
\end{equation}
with the rest of the gauge field $A'^{e^3}_{\mu_{\rm (I,II)}}(x)$ bearing the following form in the momentum space:
\begin{equation}
\begin{split}
\tilde A'^{e^3}_{\mu_{\rm (I,II)}}=&-\frac{e^3}{8}\Big[\tilde a_\mu(k)\Big(\big(\tilde a(p)\theta q\big)\big(\tilde a(q)\theta k\big)M_{1_{\rm (I,II)}}+\big(\tilde a(p)\theta\tilde a(q)\big)(q\theta k)M_{2_{\rm (I,II)}}
\\&+\big(\tilde a(p)\theta k\big)\big(\tilde a(q)\theta k\big)M'_{3_{\rm (I,II)}}\Big)
+k_\mu\Big(\big(\tilde a(p)\theta\tilde a(q)\big)\big(q\theta \tilde a(k)\big)M'_{4_{\rm (I,II)}}\Big)\Big].
\end{split}
\label{3.17}
\end{equation}
The above gauge parameters $\Xi^{e^3}_{\rm (I,II)}(x)$ have been found explicitly for both cases, the (I) and the (II),
\begin{eqnarray}
{\Xi^{e^3}_{\rm (I)}}(x)&=&-\frac{e^3}{8}\theta^{ij}\theta^{kl}\Big(2\left[a_l\partial_j a_k a_i\right]_{\star_{3'}}+\left[a_i\partial_j a_k a_l\right]_{\star_{3'}}\Big)+\mathcal O(e^4),
\label{3.15}\\
\Xi^{e^3}_{\rm (II)}(x)&=&-\frac{e^3}{4}\theta^{ij}\theta^{kl}\Big(2\left(a_i\star_2 \partial_j a_k\right)\star_2 a_l+a_i\star_2\left(\partial_j a_k\star_2 a_l\right)\Big)+\mathcal O(e^4),
\label{3.16}
\end{eqnarray}
respectively. It also turns out that the coefficients  $M'_{4_{\rm (I,II)}}$ satisfy:
\begin{equation}
M'_{4_{\rm (I,II)}}=-M_{1_{\rm (I,II)}}.
\label{3.172}
\end{equation}
Meanwhile, the $p\leftrightarrow q$ permutation symmetry of the $(\tilde a(p)\theta k)(\tilde a(q)\theta k)$ term, leads to $M'_{3_{\rm (I,II)}}$:
\begin{equation}
2M'_{3_{\rm (I,II)}}=M_{3_{\rm (I,II)}}[p,q,k]+M_{3_{\rm (I,II)}}[q,p,k],
\label{3.174}
\end{equation}
which further simplifies (\ref{3.17}).

Finishing all above transformations we now examine the remaining difference
$W_\mu(x)=A'^{e^3}_{\mu_{\rm (I)}}(x)-A'^{e^3}_{\mu_{\rm (II)}}(x)$
in momentum space:
\begin{equation}
\begin{split}
\tilde W_\mu=\tilde A'^{e^3}_{\mu_{\rm (I)}}-\tilde A'^{e^3}_{\mu_{\rm (II)}}=&e^3\Big[\tilde a_\mu(k)\Big(\big(\tilde a(p)\theta q\big)\big(\tilde a(q)\theta k\big)\tilde W_1-\frac{1}{2}\big(\tilde a(p)\theta\tilde a(q)\big)(q\theta k)\tilde W_1
\\&+\big(\tilde a(p)\theta k\big)\big(\tilde a(q)\theta k\big)\tilde W_3\Big)
-k_\mu\Big(\big(\tilde a(p)\theta\tilde a(q)\big)\big(q\theta \tilde a(k)\big)\Big)\tilde W_1\Big],
\end{split}
\label{3.18}
\end{equation}
where
\begin{eqnarray}
\tilde W_1=\frac{p\theta k}{2}\bigg[\bigg(\frac{p\theta q}{2}-\frac{p\theta k}{2}+\frac{q\theta k}{2}\bigg) f_1\bigg(\frac{p\theta q}{2},\frac{p\theta k}{2},\frac{q\theta k}{2}\bigg)
+\frac{p\theta q}{2}\frac{p\theta k}{2}\frac{q\theta k}{2} f_2\bigg(\frac{p\theta q}{2},\frac{p\theta k}{2},\frac{q\theta k}{2};0\bigg)\bigg]
=2\frac{p\theta k}{q\theta p} \tilde W_3,
\label{3.20}
\end{eqnarray}
with functions $f_{1}(a,b,c)$ and $f_{2}(a,b,c;n)$ having relatively complicated structures
\begin{equation}
\begin{split}
f_1\left(a,b,c\right)=&\frac{1}{4}\bigg(\frac{\cos(a+b+c)}{(a+b)(b+c)(a+c)(a+b+c)}+\frac{\cos(a+b-c)}{(a+b)(b+c)(a-c)(a+b-c)}
\\&-\frac{\cos(a-b+c)}{(a-b)(b-c)(a+c)(a-b+c)}-\frac{\cos(a-b-c)}{(a-b)(b+c)(a-c)(a-b-c)}
\\&-\frac{8}{(a+b+c)(a+b-c)(a-b+c)(a-b-c)}\bigg),
\end{split}
\label{3.21}
\end{equation}
\begin{equation}
f_2\left(a,b,c;n\right)= \frac{a^{2n}\cos a \,\sin b \,\sin c}{bc(a^2-b^2)(c^2-a^2)} +\frac{b^{2n}\sin a\, \cos b \,\sin c}{ac(a^2-b^2)(b^2-c^2)}+\frac{c^{2n}\sin a\, \sin b\, \cos c}{ab(b^2-c^2)(c^2-a^2)}.
\label{3.22}
\end{equation}
One can immediately observe that functions $f_1$ and $f_2$ are both completely symmetric under any permutation over a,b,c. This enables us to express the relevant part $W_\mu(x)$ of the difference between two $\theta$-exact SW maps (I) and (II) via two new generalized entirely symmetric 3-products $\diamond_1$ and $\diamond_2(n)$:
\begin{gather}
[fgh]_{\diamond_1}(x)=\int\,e^{-i(p+q+k)x}\tilde f(p)\tilde g(q)\tilde h(k)f_1\left(\frac{p\theta q}{2},\frac{p\theta k}{2},\frac{q\theta k}{2}\right),
\label{3.23}\\
[fgh]_{\diamond_2(n)}(x)=\int\,e^{-i(p+q+k)x}\tilde f(p)\tilde g(q)\tilde h(k)f_2\left(\frac{p\theta q}{2},\frac{p\theta k}{2},\frac{q\theta k}{2};n\right).
\label{3.24}
\end{gather}
After reformulating the inverse fourier transformation of (\ref{3.18}) in terms of $\diamond_1$ and $\diamond_{2(0)}$ products and some lengthy rearrangement of the fields and (the rest of) the indices we find following expression for $W_\mu(x)$ in terms of $\diamond_1$ and $\diamond_{2(0)}$ 3-products:
\begin{equation}
\begin{split}
W&_\mu(x)=-\frac{e^3}{8}\theta^{ij}\theta^{kl}\theta^{pq}\theta^{rs}\bigg([\partial_r f_{ip}f_{jk}\partial_s\partial_q f_{l\mu}]_{\diamond_1}+[\partial_r f_{ip}f_{jk}\partial_s\partial_l f_{q\mu}]_{\diamond_1}
+[\partial_p f_{ri}\partial_q f_{jk}\partial_s f_{l\mu}]_{\diamond_1}
\\&+\frac{1}{4}\theta^{ab}\theta^{cd}\left[\partial_p\partial_a f_{ri}\partial_q\partial_c f_{jk}\partial_s\partial_b\partial_d f_{l\mu}\right]_{\diamond_2(0)}+\partial_\mu\Big([\partial_p f_{ir}\partial_q f_{jk}\partial_s a_l]_{\diamond_1}
+2[\partial_p\partial_r a_i\partial_q\partial_j a_k\partial_s a_l]_{\diamond_1}
\\&-\frac{1}{12}\theta^{ab}\theta^{cd}\big(3[\partial_p\partial_a\partial_r a_i\partial_q\partial_c\partial_j a_k\partial_s\partial_b\partial_d a_l]_{\diamond_2(0)}
-[\partial_p\partial_a\partial_i a_r\partial_q\partial_c\partial_k a_j\partial_s\partial_b\partial_d a_l]_{\diamond_2(0)}\big)\Big)\bigg).
\end{split}
\label{Wmu}
\end{equation}
Note that both $f_1(a,b,c)$ and $f_2(a,b,c,n)$ have smooth $\theta\to 0$ limit as expected. Consequently $W_\mu(x)$ will end at the
$\theta^6$ order if we expand 3-products $\diamond_1$ and $\diamond_{2(0)}$ to their lowest order only. Finally, we add $\big(\partial_\mu\Xi^{e^3}_{\rm (I)}(x)-\partial_\mu\Xi^{e^3}_{\rm (II)}(x)\big)$ back to the $W_{\mu}(x)$ and obtain explicit solutions to the equations (\ref{3.6}):
\begin{gather}
\begin{split}
X^{e^3}_\mu(x)=&-\frac{e^3}{8}\theta^{ij}\theta^{kl}\theta^{pq}\theta^{rs}\Big([\partial_r f_{ip}f_{jk}\partial_s\partial_q f_{l\mu}]_{\diamond_1}+[\partial_r f_{ip}f_{jk}\partial_s\partial_l f_{q\mu}]_{\diamond_1}
+[\partial_p f_{ri}\partial_q f_{jk}\partial_s f_{l\mu}]_{\diamond_1}
\\&+\frac{1}{4}\theta^{ab}\theta^{cd}\left[\partial_p\partial_a f_{ri}\partial_q\partial_c f_{jk}\partial_s\partial_b\partial_d f_{l\mu}\right]_{\diamond_2(0)}\Big),
\end{split}
\label{3.25}\\
\begin{split}
Y^{e^3}(x)=&\frac{e^3}{8}\theta^{ij}\theta^{kl}\Big(2\left[a_l\partial_j a_k a_i\right]_{\star_{3'}}+\left[a_i\partial_j a_k a_l\right]_{\star_{3'}}-4\left(a_i\star_2 \partial_j a_k\right)\star_2 a_l-2a_i\star_2\left(\partial_j a_k\star_2 a_l\right)\Big)
\\&-\frac{e^3}{8}\theta^{ij}\theta^{kl}\theta^{pq}\theta^{rs}\Big([\partial_p f_{ir}\partial_q f_{jk}\partial_s a_l]_{\diamond_1}
+2[\partial_p\partial_r a_i\partial_q\partial_j a_k\partial_s a_l]_{\diamond_1}
\\&-\frac{1}{12}\theta^{ab}\theta^{cd}\big(3[\partial_p\partial_a\partial_r a_i\partial_q\partial_c\partial_j a_k\partial_s\partial_b\partial_d a_l]_{\diamond_2(0)}
-[\partial_p\partial_a\partial_i a_r\partial_q\partial_c\partial_k a_j\partial_s\partial_b\partial_d a_l]_{\diamond_2(0)}\big)\Big).
\end{split}
\label{3.26}
\end{gather}

The simple structure for the ambiguity between two gauge field SW maps at the $e^3$ order~\eqref{3.6} leads to the following result for the gauge field strength difference/comparison
\begin{equation}
\begin{split}
F^{e^3}_{\mu\nu_{\rm (I)}}(x)-F^{e^3}_{\mu\nu_{\rm (II)}}(x)=&\partial_\mu X^{e^3}_\nu(x)-\partial_\nu X^{e^3}_\mu(x)
\\=&\frac{e^3}{4}\theta^{ij}\theta^{kl}\theta^{pq}\theta^{rs}\bigg(\left[\partial_r\partial_p f_{\mu i}\partial_ q f_{jk}\partial_s f_{\nu l}\right]_{\diamond_1}-\left[\partial_r\partial_p f_{\mu i}f_{jk}\partial_s\partial_q f_{\nu l}\right]_{\diamond_1}+\left[\partial_p f_{\mu i}\partial_r f_{jk}\partial_q\partial_s\partial_q f_{\nu l}\right]_{\diamond_1}
\\&+\left[\partial_r f_{pi}\partial_s\partial_j f_{\mu k}\partial_s\partial_q f_{\nu l}\right]_{\diamond_1}-\left[\partial_r f_{pi}\partial_j f_{\mu k}\partial_s\partial_q f_{\nu l}\right]_{\diamond_1}+\left[f_{p i}\partial_r\partial_j f_{\mu k}\partial_s\partial_q f_{\nu l}\right]_{\diamond_1}
\\&+\frac{1}{4}\theta^{ab}\theta^{cd}\left(\left[\partial_p\partial_r\partial_a f_{\mu i}\partial_q\partial_c f_{jk}\partial_s\partial_b\partial_d f_{\nu l}\right]_{\diamond_{2(0)}}+\left[\partial_p\partial_a f_{ri}\partial_q\partial_c\partial_j f_{\mu k}\partial_s\partial_b\partial_d f_{\nu l}\right]_{\diamond_{2(0)}}\right)\bigg)
\\&+\frac{e^3}{8}\theta^{ij}\theta^{kl}\theta^{pq}\theta^{rs}\Big(2\left[\partial_r f_{pi} f_{jk}\partial_s\partial_q\partial_l f_{\mu\nu}\right]_{\diamond_1}-\left[\partial_r f_{pi}\partial_s f_{jk}\partial_q\partial_l f_{\mu\nu}\right]_{\diamond_1}
\\&-\frac{1}{4}\theta^{ab}\theta^{cd}\left[\partial_r\partial_a f_{pi}\partial_s\partial_c f_{jk}\partial_q\partial_b\partial_d\partial_l f_{\mu\nu}\right]_{\diamond_{2(0)}}\Big).
\end{split}
\label{3.27}
\end{equation}
Clearly the $Y^{e^3}(x)$ related terms drop out, leaving (\ref{3.27}) containing only the $\rm U(1)$ gauge field strengths.

\section{The $\theta$-exact gauge field strength up to the $e^3$ order}

In the last sections we examined the SW map ambiguities between two known gauge field expansions. This section focus on another type of freedom within each of the two field strength SW map expansions. Our motivation is to find certain $\theta$-exact alternative of the earlier $\theta$-iterative freedom parameters. In past majority of the studies on SW map ambiguities follows the $\theta$-iterative field redefinition procedure in~\cite{Bichl:2001cq,Asakawa:1999cu}. At $\theta^1$ order this procedure introduces the following correction to the gauge field expansion~\cite{Buric:2006wm}
\begin{equation}
\varphi_\mu=\frac{b}{4} e\theta^{ij}D_\mu f_{ij},
\label{5.1}
\end{equation}
which then gives the following gauge field strength correction,
\begin{equation}
\Phi_{\mu\nu}=D_\mu\varphi_\nu-D_\nu\varphi_\mu=\frac{b}{4} e\theta^{ij}f_{\mu\nu}f_{ij},
\label{5.2}
\end{equation}
and consequently modifies the action into
\begin{equation}
S^{e^2\theta^1}_b=-\int e^2\theta^{ij}f^{\mu\nu}\left(f_{\mu i} f_{\nu j}-\frac{1+b}{4}f_{ij} f_{\mu\nu}\right).
\label{5.4}
\end{equation}

To find a $\theta$-exact alternative of \eqref{5.4} we first study the $e^2$ order field strength SW map expansion. Gauge field expansions $A_{\mu_{\rm (I)}}(x)$ and $A_{\mu_{\rm (II)}}(x)$, from (\ref{2.10}) and  (\ref{2.12}) respectively, lead to the same noncommutative $\rm U_\star(1)$ gauge field strength expansion up to the $e^2$ order
\begin{equation}
\begin{split}
F_{\mu\nu}(x)=&e f_{\mu\nu}
+ e^2\theta^{ij}
\Big(f_{\mu i}\star_{2}f_{\nu j}-a_i\star_2\partial_j
f_{\mu\nu}\Big)+\mathcal{O}(e^3)\,.
\end{split}
\label{e2}
\end{equation}
On the other hand, at the $e^2$ order, the general consistency condition for the gauge field strength
\begin{equation}
\delta_\lambda F_{\mu\nu}=i\left[\Lambda\stackrel{\star}{,} F_{\mu\nu}\right],
\label{de2}
\end{equation}
becomes
\begin{equation}
\delta_\lambda F^{e^2}_{\mu\nu}=ie^2\left[\lambda\stackrel{\star}{,} f_{\mu\nu}\right].
\label{ce2}
\end{equation}
Examining the gauge field strength \eqref{e2} we find that the variation of the first term at $e^2$ order, $e^2\theta^{ij}
f_{\mu i}\star_{2}f_{\nu j}$, vanishes, therefore the consistency condition \eqref{ce2} is fulfilled solely through the second term $-e^2\theta^{ij}a_i\star_2\partial_j f_{\mu\nu}$ as
\begin{equation}
\delta_\lambda \left(-e^2\theta^{ij}a_i\star_2\partial_j f_{\mu\nu}\right)=-e^2\theta^{ij}\partial_i\lambda\star_2\partial_j f_{\mu\nu}=ie^2[\lambda\stackrel{\star}{,}f_{\mu\nu}],
\label{59}
\end{equation}
thanking to the relation between the $\star_2$-product and the $\star$-commutator (\ref{fe2}). This observation promotes us to put an arbitrary parameter $\kappa$ in front of the term $e^2\theta^{ij}f_{\mu i}\star_2 f_{\nu j}$ in equation \eqref{e2} since this does not break the $e^2$ order consistency condition \eqref{ce2}. Such a procedure leads to
the $\kappa$-deformed gauge field strength up to the $e^2$ order
\begin{equation}
\begin{split}
F_{\mu\nu}(x)_{\kappa}=&e f_{\mu\nu}
+ e^2\theta^{ij}
\Big(\kappa f_{\mu i}\star_{2}f_{\nu j}-a_i\star_2\partial_j
f_{\mu\nu}\Big)+\mathcal{O}(e^3)\,.
\end{split}
\label{e2k}
\end{equation}
Restriction of the $F_{\mu\nu}(x)_{\kappa}$ to the $\theta^1$ order gives
\begin{equation}
\begin{split}
F_{\mu\nu}(x)_{\kappa}^{\theta^1}=&e^2\theta^{ij}
\Big(\kappa f_{\mu i}f_{\nu j}-a_i\partial_j
f_{\mu\nu}\Big)\,.
\end{split}
\label{e2k}
\end{equation}
Also the deformed action at the $\theta^1$ order reads
\begin{equation}
S^{e^2\theta^1}_{\kappa}=-\int e^2\theta^{ij}f^{\mu\nu}\left(\kappa f_{\mu i} f_{\nu j}-\frac{1}{4}f_{ij} f_{\mu\nu}\right).
\label{5.3}
\end{equation}
Here we see that the $b$ correction \eqref{5.2} to the gauge field strength does not match the $\kappa$ correction in (\ref{e2k}). However the $b$ and $\kappa$ corrections are instead connected by the (inter-)actions since the $\kappa$ and/or $a=1+b$ present the ratio between two gauge invariant terms in an inverted fashion. For this reason we consider $\kappa$ deformation as a possible substitute for the $b$ ($a$ in literatures) modification in the $\theta$-exact approach.

To extend the $\kappa$ deformation to $e^3$ order we must handle the effect of $\kappa$ in the $e^3$ order consistency relation as well as the identification of possible new gauge invariant terms. This can be done by solving the consistency relation
\begin{equation}
\begin{split}
\delta_\lambda F^{e^3}_{\mu\nu}(x)_{\kappa}=&i e\left(\Big[\Lambda^{e^2}\stackrel{\star}{,}f_{\mu\nu}\Big]+\Big[\lambda\stackrel{\star}{,}F^{e^2}_{\mu\nu}(x)_{\kappa}\Big]\right)
\\=&i e\left(\Big[\Lambda^{e^2}\stackrel{\star}{,}f_{\mu\nu}\Big]+\Big[\lambda\stackrel{\star}{,}e^2\theta^{ij}
\big(\kappa f_{\mu i}\star_{2}f_{\nu j}-a_i\star_2\partial_j
f_{\mu\nu}\big)\Big]\right).
\end{split}
\label{e3k}
\end{equation}
We start by observing that both SW maps in section II satisfy the $e^3$ order consistency relation \eqref{e3k} when $\kappa=1$, i.e. without the $\kappa$-deformation; then identify, within the un\-deformed noncommutative field strength $F_{\mu\nu_{\rm (I,II)}}(x)$, those terms relevant to the to-be-deformed term $i\theta^{ij}\left[\lambda\stackrel{\star}{,}f_{\mu i}\star_2 f_{\nu j}\right]$ and make them $\kappa$-proportional. We also search for possible $\kappa$-unrelated freedom/ambiguity in the un\-deformed noncommutative field strength.

\subsection{Gauge field strength from the Seiberg-Witten map (I)}

The easiest way to determine the gauge field strength corresponding to the gauge field
$A_{\mu_{\rm (I)}}(x)$ is by solving directly the Seiberg-Witten differential equation for the gauge field strength \cite{Seiberg:1999vs}
\begin{equation}
\frac{d}{dt} F_{\mu\nu_{\rm (I)}}(x)=\frac{1}{4}\theta^{ij}\bigg[2\Big\{F_{\mu i}\stackrel{\star_t}{,}F_{\nu j}\Big\}-\Big\{A_i\stackrel{\star_t}{,}\Big(2\partial_jF_{\mu\nu}-i\left[A_j\stackrel{\star_t}{,}F_{\mu\nu}\right]\Big)\Big\}\bigg],
\label{4.5}
\end{equation}
which, at the $e^3$ order yields
\begin{equation}
\begin{split}
F^{e^3}_{\mu\nu_{\rm (I)}}(x)
&=\frac{e^3}{2}\theta^{ij}\theta^{kl}\bigg[\left(\left[f_{\mu k}f_{\nu i} f_{l j}\right]_{\star_{3'}}+\left[f_{\nu l}f_{\mu i}f_{kj}\right]_{\star_{3'}}\right)
-\left(\left[f_{\nu l}a_i\partial_j f_{\mu k}\right]_{\star_{3'}}+\left[f_{\mu k}a_i\partial_j f_{\nu l}\right]_{\star_{3'}}
+\left[a_k\partial_l\left(f_{\mu i}f_{\nu j}\right)\right]_{\star_{3'}}\right)
\\&
+\left[a_i\partial_j a_k \partial_l f_{\mu\nu}\right]_{\star_{3'}}+\left[\partial_l f_{\mu\nu}a_i\partial_j a_k\right]_{\star_{3'}}
+\left[a_k a_i \partial_l\partial_j f_{\mu\nu}\right]_{\star_{3'}} -\frac{1}{2}\left(\left[a_i\partial_k a_j\partial_l f_{\mu\nu}\right]_{\star_{3'}}+\left[\partial_l f_{\mu\nu}a_i\partial_k a_j\right]_{\star_{3'}}\right)\bigg].
\end{split}
\label{4.6}
\end{equation}

Here we notice facts. First is that among all above terms in the equation (\ref{4.6}), the first two in the first line are manifestly invariant under the commutative gauge transformation and antisymmetric under the $\mu\leftrightarrow\nu$ permutation, therefore could be a subject to the free variation, i.e. associated with new deformation (weight) parameter $\kappa_1$.

Next considering the next three terms in the first line of (\ref{4.6}), with the help of \eqref{fe3p}
we find out that the sum of these three terms together satisfy the following transformation property
\begin{equation}
\delta_\lambda\frac{1}{2}\theta^{ij}\theta^{kl}\left(\left[f_{\nu l}a_i\partial_j f_{\mu k}\right]_{\star_{3'}}+\left[f_{\mu k}a_i\partial_j f_{\nu l}\right]_{\star_{3'}}+\left[a_k\partial_l\left(f_{\mu i}f_{\nu j}\right)\right]_{\star_{3'}}\right)=-i\theta^{kl}\left[\lambda\stackrel{\star}{,}f_{\mu k}\star_2 f_{\nu l}\right].
\label{4.8}
\end{equation}
Thus, the second fact is that they are relevant subject for the $\kappa$-deformation at the $e^3$ order.

These two facts lead us to an extended $(\kappa,\kappa_1)$-deformation of the field strength (\ref{4.6}):
\begin{equation}
\begin{split}
F^{e^3}_{\mu\nu_{\rm (I)}}(x)_{\kappa,\kappa_1}&=
\frac{e^3}{2}\theta^{ij}\theta^{kl}\bigg[\kappa_1\left(\left[f_{\mu k}f_{\nu i} f_{l j}\right]_{\star_{3'}}+\left[f_{\nu l}f_{\mu i}f_{kj}\right]_{\star_{3'}}\right)
\\&-\kappa\left(\left[f_{\nu l}a_i\partial_j f_{\mu k}\right]_{\star_{3'}}
+\left[f_{\mu k}a_i\partial_j f_{\nu l}\right]_{\star_{3'}}+\left[a_k\partial_l\left(f_{\mu i}f_{\nu j}\right)\right]_{\star_{3'}}\right)
\\&+\left[a_i\partial_j a_k \partial_l f_{\mu\nu}\right]_{\star_{3'}}
+\left[\partial_l f_{\mu\nu}a_i\partial_j a_k\right]_{\star_{3'}}+\left[a_k a_i \partial_l\partial_j f_{\mu\nu}\right]_{\star_{3'}}
-\frac{1}{2}\Big(\left[a_i\partial_k a_j\partial_l f_{\mu\nu}\right]_{\star_{3'}}
+\left[\partial_l f_{\mu\nu}a_i\partial_k a_j\right]_{\star_{3'}}\Big)\bigg]\,.
\end{split}
\label{4.11}
\end{equation}

\subsection{Gauge field strength from the Seiberg-Witten map  (II)}

Next we consider the $e^3$ order $\theta$-exact gauge field strength from $A_{\mu_{\rm (II)}}(x)$, which can be expressed as below
\begin{equation}
\begin{split}
&F^{e^3}_{\mu\nu_{\rm (II)}}(x)=e^3\theta^{ij}\theta^{kl} \Big[f_{\mu i}\star_2\left(f_{jk}\star_2 f_{l\nu}\right)+f_{l\nu}\star_2\left(f_{jk}\star_2 f_{\mu i}\right)-\left[f_{\mu i}f_{jk}f_{l\nu}\right]_{\star_3}
\\&-\big((a_i\star_2\partial_j f_{\mu k})\star_2 f_{\nu l}+(a_i\star_2\partial_j f_{\nu l})\star_2 f_{\mu k}-[a_i\partial_j (f_{\mu k}f_{\nu l})]_{\star_3}\big)
\\&- a_i\star_2\partial_j\left(f_{\mu k}\star_2 f_{\nu l}\right)+(a_i\star_2\partial_j a_k)\star_2\partial_l f_{\mu\nu}
\\&+a_i\star_2(\partial_j a_k\star_2\partial_l f_{\mu\nu})+a_i\star_2(a_k\star_2\partial_j\partial_l f_{\mu\nu})-[a_i\partial_j a_k\partial_l f_{\mu\nu}]_{\star_{3}}
\\&-\frac{1}{2}\Big(a_i\star_2(\partial_k a_j\star_2\partial_l f_{\mu\nu})+(a_i\star_2\partial_k a_j)\star_2\partial_l f_{\mu\nu}-[a_i\partial_k a_j\partial_l f_{\mu\nu}]_{\star_3}
+[a_ia_k\partial_j\partial_l f_{\mu\nu}]_{\star_3}\Big)\Big].
\label{4.1}
\end{split}
\end{equation}
Using the basic relation \eqref{fe3}
we can show that the infinitesimal commutative gauge transformation of the parentheses in the second line of (\ref{4.1})
\begin{equation}
\begin{split}
&\delta_\lambda\theta^{ij}\theta^{kl}\big((a_i\star_2\partial_j f_{\mu k})\star_2 f_{\nu l}+(a_i\star_2\partial_j f_{\nu l})\star_2 f_{\mu k}-[a_i\partial_j (f_{\mu k}f_{\nu l})]_{\star_3}\big)
\\&=\theta^{ij}\theta^{kl}(\partial_i\lambda\star_2\partial_j f_{\mu k})\star_2 f_{\nu l}+(\partial_i\lambda\star_2\partial_j f_{\nu l})\star_2 f_{\mu k}-[\partial_i\lambda\partial_j (f_{\mu k}f_{\nu l})]_{\star_3}
\\&=i\left([\lambda\stackrel{\star}{,}f_{\mu k}]\star_2 f_{\nu l}+[\lambda\stackrel{\star}{,}f_{\nu l}]\star_2 f_{\mu k}+f_{\nu l}\star_2[f_{\mu k}\stackrel{\star}{,}\lambda]+f_{\mu k}\star_2[f_{\nu l}\stackrel{\star}{,}\lambda]\right)=0
\end{split}
\label{69}
\end{equation}
vanishes. One can further turn this parentheses into a manifestly gauge invariant form with the help of
the 3-products $\diamond_{2(n)}$:
\begin{equation}
\begin{split}
&\theta^{ij}\theta^{kl}\Big((a_i\star_2\partial_j f_{\mu k})\star_2 f_{\nu l}+(a_i\star_2\partial_j f_{\nu l})\star_2 f_{\mu k}-[a_i\partial_j (f_{\mu k}f_{\nu l})]_{\star_3}\Big)=\frac{1}{4}\theta^{ij}\theta^{kl}\theta^{pq}\theta^{rs}\Big(\left[f_{pi}\partial_j\partial_r f_{\mu k}\partial_q\partial_s f_{\nu l}\right]_{\diamond_2(1)}
\\&+\frac{1}{4}\theta^{ab}\theta^{cd}\left[\partial_p\partial_a f_{ri}\partial_q\partial_c\partial_j f_{\mu k}\partial_s\partial_b\partial_d f_{\nu l}+\partial_p f_{ir}\partial_q\partial_a\partial_c\partial_j f_{\mu k}\partial_s\partial_b\partial_d f_{\nu l}+\partial_p f_{ir}\partial_q\partial_a\partial_c\partial_j f_{\nu l}\partial_s\partial_b\partial_d f_{\mu k}\right]_{\diamond_2(0)}\Big),
\label{70}
\end{split}
\end{equation}
therefore we conclude that that the first two lines in (\ref{4.1}) do not contribute to $\delta_\lambda F^{e^3}_{\mu\nu}$. Among the rest of the terms, we notice that the first one is compatible with the $\star_2$-commutator, since:
\begin{equation}
\delta_\lambda\big(- \theta^{ij}\theta^{kl}a_i\star_2\partial_j\left(f_{\mu k}\star_2 f_{\nu l}\right)\big)=i\theta^{kl}\left[\lambda\stackrel{\star}{,}f_{\mu k}\star_2 f_{\nu l}\right].
\label{4.2}
\end{equation}
Thus, this term alone gives the formal NC transformation of the fully commutative gauge field strength term $\theta^{ij}f_{\mu i}\star_2 f_{\nu j}$ at $e^3$ order. Therefore multiplying equation \eqref{4.2} by the $\kappa$ parameter ensures the compatibility at the $e^3$ order.

It is also straightforward to notice that two more additional free variation could be performed on $F^{e^3}_{\mu\nu_{\rm (II)}}(x)$ via multiplication of the manifestly gauge invariant first two lines of equation \eqref{4.1} by two new deformation parameter $\kappa'_1$ and $\kappa'_2$, respectively. This way we obtain the the ($\kappa,\kappa'_1,\kappa'_2$)-deformed extension for the gauge field strength at the $F^{e^3}_{{\mu\nu}_{\rm (II)}}(x)$
\begin{equation}
\begin{split}
&F^{e^3}_{\mu\nu_{\rm (II)}}(x)_{\kappa,\kappa'_1,\kappa'_2}=e^3\theta^{ij}\theta^{kl}\Bigg[\kappa'_1\Big(f_{\mu i}\star_2\left(f_{jk}\star_2 f_{l\nu}\right)+f_{l\nu}\star_2\left(f_{jk}\star_2 f_{\mu i}\right)-\left[f_{\mu i}f_{jk}f_{l\nu}\right]_{\star_3}\Big)
\\&-\kappa'_2\frac{1}{4}\theta^{pq}\theta^{rs}\Big(\left[f_{pi}\partial_j\partial_r f_{\mu k}\partial_q\partial_s f_{\nu l}\right]_{\diamond_2(1)}
+\frac{1}{4}\theta^{ab}\theta^{cd}\left[\partial_p\partial_a f_{ri}\partial_q\partial_c\partial_j f_{\mu k}\partial_s\partial_b\partial_d f_{\nu l}
\right.\\&\left.+\partial_p f_{ir}\partial_q\partial_a\partial_c\partial_j f_{\mu k}\partial_s\partial_b\partial_d f_{\nu l}+\partial_p f_{ir}\partial_q\partial_a\partial_c\partial_j f_{\nu l}\partial_s\partial_b\partial_d f_{\mu k}\right]_{\diamond_2(0)}\Big)
\\&-\kappa a_i\star_2\partial_j\left(f_{\mu k}\star_2 f_{\nu l}\right)+(a_i\star_2\partial_j a_k)\star_2\partial_l f_{\mu\nu}
+a_i\star_2(\partial_j a_k\star_2\partial_l f_{\mu\nu})+a_i\star_2(a_k\star_2\partial_j\partial_l f_{\mu\nu})-[a_i\partial_j a_k\partial_l f_{\mu\nu}]_{\star_{3}}
\\&-\frac{1}{2}\Big(a_i\star_2(\partial_k a_j\star_2\partial_l f_{\mu\nu})+(a_i\star_2\partial_k a_j)\star_2\partial_l f_{\mu\nu}-[a_i\partial_k a_j\partial_l f_{\mu\nu}]_{\star_3}+[a_ia_k\partial_j\partial_l f_{\mu\nu}]_{\star_3}\Big)\Bigg].
\label{4.4}
\end{split}
\end{equation}

Inspired by the form of \eqref{4.4}, we add and subtract $a_i\star_2\partial_j\left(f_{\mu k}\star_2 f_{\nu l}\right)$ in \eqref{4.6}, then assign $\kappa$ to the $a_i\star_2\partial_j\left(f_{\mu k}\star_2 f_{\nu l}\right)$ term, and assign new parameter $\kappa_2$ to the sum of the terms $(\left[f_{\nu l}a_i\partial_j f_{\mu k}\right]_{\star_{3'}}
+\left[f_{\mu k}a_i\partial_j f_{\nu l}\right]_{\star_{3'}}+\left[a_k\partial_l\left(f_{\mu i}f_{\nu j}\right)\right]_{\star_{3'}}-2a_i\star_2\partial_j\left(f_{\mu k}\star_2 f_{\nu l}\right))$. This way  \eqref{4.11} is generalized into essentially the same form as the equation (\ref{4.4}). This leads to
\begin{equation}
\begin{split}
F^{e^3}_{\mu\nu_{\rm (I)}}(x)_{\kappa,\kappa_1,\kappa_2}&=
\frac{e^3}{2}\theta^{ij}\theta^{kl}\bigg[\kappa_1\left(\left[f_{\mu k}f_{\nu i} f_{l j}\right]_{\star_{3'}}+\left[f_{\nu l}f_{\mu i}f_{kj}\right]_{\star_{3'}}\right)-\kappa a_i\star_2\partial_j\left(f_{\mu k}\star_2 f_{\nu l}\right)
\\&-\kappa_2\left(\left[f_{\nu l}a_i\partial_j f_{\mu k}\right]_{\star_{3'}}
+\left[f_{\mu k}a_i\partial_j f_{\nu l}\right]_{\star_{3'}}+\left[a_k\partial_l\left(f_{\mu i}f_{\nu j}\right)\right]_{\star_{3'}}-2a_i\star_2\partial_j\left(f_{\mu k}\star_2 f_{\nu l}\right)\right)
\\&+\left[a_i\partial_j a_k \partial_l f_{\mu\nu}\right]_{\star_{3'}}
+\left[\partial_l f_{\mu\nu}a_i\partial_j a_k\right]_{\star_{3'}}+\left[a_k a_i \partial_l\partial_j f_{\mu\nu}\right]_{\star_{3'}}
-\frac{1}{2}\Big(\left[a_i\partial_k a_j\partial_l f_{\mu\nu}\right]_{\star_{3'}}
+\left[\partial_l f_{\mu\nu}a_i\partial_k a_j\right]_{\star_{3'}}\Big)\bigg]\,.
\label{72}
\end{split}
\end{equation}
Note that the $\kappa_2$ proportional part can be expressed  as follows using diamond $\diamond_1$ and $\diamond_{2(0,1)}$ products,
\begin{equation}
\begin{split}
&\left(\left[f_{\nu l}a_i\partial_j f_{\mu k}\right]_{\star_{3'}}
+\left[f_{\mu k}a_i\partial_j f_{\nu l}\right]_{\star_{3'}}+\left[a_k\partial_l\left(f_{\mu i}f_{\nu j}\right)\right]_{\star_{3'}}-2a_i\star_2\partial_j\left(f_{\mu k}\star_2 f_{\nu l}\right)\right)
\\&=\frac{e^3}{4}\theta^{ij}\theta^{kl}\theta^{pq}\theta^{rs}\bigg(\left[f_{pi}\partial_j\partial_r f_{\mu k}\partial_q\partial_s f_{\nu l}\right]_{\diamond_2(1)}-\left[\partial_r f_{pi}\partial_s\partial_j f_{\mu k}\partial_s\partial_q f_{\nu l}\right]_{\diamond_1}+\left[\partial_r f_{pi}\partial_j f_{\mu k}\partial_s\partial_q f_{\nu l}\right]_{\diamond_1}
\\&-\left[f_{p i}\partial_r\partial_j f_{\mu k}\partial_s\partial_q f_{\nu l}\right]_{\diamond_1}
+\frac{1}{4}\theta^{ab}\theta^{cd}\left(\left[\partial_p f_{ir}\partial_q\partial_a\partial_c\partial_j f_{\mu k}\partial_s\partial_b\partial_d f_{\nu l}+\partial_p f_{ir}\partial_q\partial_a\partial_c\partial_j f_{\nu l}\partial_s\partial_b\partial_d f_{\mu k}\right]_{\diamond_2(0)}\right)\bigg).
\end{split}
\label{tkg2}
\end{equation}
Using \eqref{3.20} one can show that difference between first two terms in (\ref{4.6}) and the first line in (\ref{4.1}) becomes
\begin{equation}
\begin{split}
&\frac{1}{2}\left(\left[f_{\mu k}f_{\nu i} f_{l j}\right]_{\star_{3'}}+\left[f_{\nu l}f_{\mu i}f_{kj}\right]_{\star_{3'}}\right)-\Big(f_{\mu i}\star_2\left(f_{jk}\star_2 f_{l\nu}\right)+f_{l\nu}\star_2\left(f_{jk}\star_2 f_{\mu i}\right)-\left[f_{\mu i}f_{jk}f_{l\nu}\right]_{\star_3}\Big)
\\&=\frac{1}{4}\theta^{ij}\theta^{kl}\theta^{pq}\theta^{rs}\Big(\left[\partial_r\partial_p f_{\mu i}\partial_ q f_{jk}\partial_s f_{\nu l}\right]_{\diamond_1}-\left[\partial_r\partial_p f_{\mu i}f_{jk}\partial_s\partial_q f_{\nu l}\right]_{\diamond_1}+\left[\partial_p f_{\mu i}\partial_r f_{jk}\partial_q\partial_s\partial_q f_{\nu l}\right]_{\diamond_1}
\\&+\frac{1}{4}\theta^{ab}\theta^{cd}\left[\partial_p\partial_r\partial_a f_{\mu i}\partial_q\partial_c f_{jk}\partial_s\partial_b\partial_d f_{\nu l}\right]_{\diamond_{2(0)}}\Big).
\label{75}
\end{split}
\end{equation}
Consequently the difference between equations (\ref{72}) and (\ref{4.4}) in the case without $\kappa$, $\kappa_i$ and $\kappa'_i$-deformations, that is for $\kappa=\kappa_{i=1,2}=\kappa'_{i=1,2}=1$, gives exactly equation (\ref{3.27})\footnote{The rest of the terms in (\ref{4.11}) and (\ref{4.4}) receive no deformation, they all contain $f_{\mu\nu}$ therefore arise when the partial derivative $\partial_{\mu(\nu)}$ hits the gauge field carrying external index $\nu(\mu)$, respectively. One can show that they equal to the terms containing $f_{\mu\nu}$ in (\ref{3.27}) following the procedure exactly identical to the section III.}, proving consistency of our computations as it should.

\section{Discussion and conclusion}

In this article we study the $e^3$ order $\theta$-exact Seiberg-Witten map expansion of $\rm U_\star(1)$ gauge field theory following an important recent progress in solving the $\theta$-exact SW map expansions for arbitrary gauge groups/representations \cite{Martin:2012aw,Martin:2015nna}. We first focus on the ambiguities between two distinct $\theta$-exact SW map expansions: First expansion (I) is obtained by solving Seiberg-Witten differential equations $\theta$-exactly \cite{Seiberg:1999vs,Martin:2012aw}, the other (II) by inverting a known SW solution~\cite{Mehen:2000vs}. Since the maps relate NC gauge orbits with ordinary ones, there are two types of freedoms: generic redefinitions or gauge transformations of the NC gauge fields, only the former will contribute to the dynamics. The redefinition freedom should be, and it was, taken into account when dealing with pathologies in the photon and neutrino two-point functions \cite{Horvat:2013rga,Trampetic:2014dea}. We then use/apply these two SW map expansions up to the $e^3$ order to study the corresponding field strength expansions and discuss the gauge inspired freedom/deformation parameters in each of the field strength expansions.

In the first part of the study, we manage to determine the ambiguity between these two maps explicitly and subject it into the standard form given in~\cite{Barnich:2002pb,Barnich:2003wq} and show that the difference of these two SW maps  for the gauge field at the $e^{3}$ order (\ref{3.27}) is generic instead of not just gauge transformations.  We find that SW map ambiguity between these two $\rm U_\star(1)$ gauge field expansions is decomposed into totally symmetric momentum structures which have lowest order at the $\theta^4$ or $\theta^6$, which means that the the ambiguity can be expressed in terms of several commutative (yet non-associative) 3-products. This crucial decomposition enables us to perform permutation and relabeling within the commutative 3-products and obtain the $\theta$-exact expression for ambiguity in terms of commutative field strength $f_{\mu\nu}$, as shown in the section III. Our observation thus indicates that at the $e^3$ order (even just) the $\rm U_\star(1)$ SW map expansion can possess much profounder structures than the prior order.

The next stage we extend the $e^2$ order gauge field strength deformation parameter $\kappa$~\cite{Trampetic:2014dea} to the $e^3$ order. We identify that part of the $e^3$ order gauge field strength should be multiplied by $\kappa$ to keep the consistency condition, while there are other parts which are invariant under the commutative gauge transformations by themselves, thus each of them can be varied independently like the $\kappa$ proportional part in the $e^2$ order. This promotes the introduction, alongside $\kappa$, new parameters $\kappa_{1,2}$ and $\kappa'_{1,2}$ for maps (I) and (II) respectively. Each pair of $\kappa(\kappa')_i$ proportional parts bear identical structure at $\theta^2$ order, yet they differ from $\theta^4$ order on. The difference between each $\kappa_i$ and $\kappa'_i$ pairs can be put into a relatively compact form using the generalized star products $\diamond_1$, $\diamond_{2(0)}$ and $\diamond_{2(1)}$ defined in section III. The total difference matches the result \eqref{3.27} derived from gauge field ambiguity when all deformation parameters are switched off.

Besides its own manifestness, results on the gauge field strength expansion in this paper can enable the construction of $\theta$-exact, and $(\kappa,\kappa_i)$ and/or $(\kappa,\kappa'_i)$ -deformed $\rm U_\star(1)$ gauge theory  (pure noncommutative Yang-Mills gauge theory action) up to the four-photon coupling term, which should then lead to the completion of the one-loop photon two-point function computation started in~\cite{Horvat:2013rga} by adding the four-photon tadpole diagram contributions. We hope that, like the $\kappa$ parameter in the bubble diagram contribution to the photon polarization tensor~\cite{Horvat:2013rga}, exploring the extended deformation freedom parameter space $(\kappa,\kappa_i)$ and/or $(\kappa,\kappa'_i)$ would provide enough control over the pathological divergences in the four-photon-tadpole diagram. The same term should also contribute to the NC phenomenology at extreme energies, for example tree-level NCQED contributions to the $2 \to 2$ scattering processes like $\gamma\gamma\to\gamma\gamma$ etc.~\cite{Hewett:2000zp,Godfrey:2001yy,Mathews:2000we,Baek:2001ty}.

Knowing the fact that prior studies based on $\star_2$-product have given rise to profound pathologies in both theory and phenomenology, even more with the presence of the gauge freedom parameter $\kappa$ ~\cite{Horvat:2011bs,Horvat:2011qg,Horvat:2012vn,Horvat:2013rga}, we positively expect that our work in this article will form a universal basis for future studies on the various potential physical effects of the generalized star products and the higher order gauge freedom parameters $\kappa_i$s and $\kappa'_i$s.

Finally it worths notice that despite its profound nature our study on the SW map ambiguity in this paper limits to only two distinct SW map expansions. There should be still many other variants available. The current methods for solving SW map(s), for example open-Wilson line operators~\cite{Mehen:2000vs}, string/D-brane inspired analysis~\cite{Seiberg:1999vs,Liu:2000ad,Liu:2000mja,Okawa:2001mv}, Batalin-Vilkovisky (BV) formalism and (generalized) Seiberg-Witten differential equations~\cite{Brace:2001fj,Brace:2001rd,Cerchiai:2002ss,Barnich:2002pb,Barnich:2003wq,Martin:2012aw,Martin:2015nna} etc., are extremely powerful in finding specific (sometimes closed form) solutions, yet normally does not provide us all possible maps simultaneously.\footnote{The Kontsevich formality based approach \cite{Jurco:2001my} does capture ambiguities in the SW map: They are related to (formal) changes of coordinates. (See the discussion in the context of the somewhat cryptic Fig. 1 in ref \cite{Jurco:2001my}.) This procedure essentially produces all possible maps, yet it is difficult to derive the explicit expansion(s) needed for practical purposes. We are particularly grateful to Peter Schupp for comments on the Kontsevich formality approach and the connection between SW maps and the Morita equivalence among star-products.} SW maps also relate Morita equivalent star products on Poisson manifolds, their non-uniqueness can be understood as a local gauge freedom in this context~\cite{Jurco:2001my,Jurco:2001kp}, which may help to understand the background (in-)dependence of the string theory. It would be delightful to see any progress alone this line in the near future.

\noindent
\acknowledgments
The work/project of Josip Trampetic is conducted under the European Commission and the Croatian Ministry of Science, Education and Sports Co-Financing Agreement No. 291823. In particular J.T. acknowledges project financing by The Marie Curie FP7-PEOPLE-2011-COFUND program NEWFELPRO: Grant Agreement No. 69.  J.T. would also like to acknowledge support of Alexander von Humboldt Foundation (KRO 1028995 STP), and  Max-Planck-Institute for Physics and W. Hollik for hospitality. We would like to acknowledge the support of the Mainz Institute for Theoretical Physics (MITP) for its hospitality, partial support, and for enabling us to complete (a significant portion of) this work. We would like to thank J. Erdmenger,  W. Hollik, A. Ilakovac, C. Martin and P. Schupp for fruitful discussions. A great deal of computation was done by using ${\rm Mathematica}$~8.0~\cite{mathematica}. Special thanks to D. Kekez for computer hard/software supports.

\end{document}